\documentclass[a4paper, 12pt]{article}

\usepackage{framed,multirow}
\usepackage{bm}
\usepackage{authblk}
\usepackage{dsfont}  %for the indicator function (font type 1)
\usepackage{amssymb}
\usepackage{amsmath}
\usepackage{latexsym}
\usepackage{subcaption}
\usepackage{multirow}
\usepackage{graphicx, changepage, mathtools, array}
\usepackage{color}
\usepackage[figuresright]{rotating}

\providecommand{\keywords}[1]{\textbf{\textit{Keywords ---}} #1}

\begin{document}
\title{Dealing with overdispersion in multivariate count data}
\author{Noemi Corsini, Cinzia Viroli \thanks{cinzia.viroli@unibo.it} \\ \small{Department of Statistical Sciences, University of Bologna} \\ \small{via Belle Arti 41, 40126, Bologna, Italy}}
\date{}
\maketitle

\begin{abstract}
The problem of overdispersion in multivariate count data is a challenging issue. Nowadays, it covers a central role mainly due to the relevance of modern technologies data, such as Next Generation Sequencing and textual data from the web or digital collections. This work presents a comprehensive analysis of the likelihood-based models for extra-variation data proposed in the scientific literature. Particular attention will be paid to the models feasible for high-dimensional data. A new approach together with its parametric-estimation procedure is proposed. It is a deeper version of the Dirichlet-Multinomial distribution and it leads to important results allowing to get a better approximation of the observed variability. A significative comparison of these models is made through two different simulation studies that both confirm that the new model considered in this work allows to achieve the best results.
\end{abstract}
\keywords{Extra-variation $\>$ Mixture models $\>$ Deep Learning $\>$ Maximum Likelihood}

\section{Introduction}
\label{sec:introduction}
The overdispersion or extra-variation is a recurring phenomenon when dealing with counts and categorical data. In particular, it often occurs that after fitting a binomial, a multinomial or a Poisson model to the data, the sampling variation is greater than the estimated variation accounted by the model. In other words, the data exhibit a larger variability than that the model is able to explain \cite{Poortema1999}. Overdispersion has specific causes and consequences. It may arise as result of the data collection and aggregation, such as  clumped sampling \cite{efron86}; it may due to correlation between individual responses or to additional experimental variability. Inferential consequences are imprecise estimates and biased standard errors that make model selection, interpretability and prediction unreliable.

In this work we focus our analysis on multivariate count data, that are becoming more recurrent thanks to recent technologies such as web scraping for textual data \cite{Munzert2015} or Next Generation Sequencing data \cite{Wang2009Jan}. In both situations, we observe multivariate count data often inflated by a large amount of zeros (words rarely used or not-expressed genes) or correlated responses. As a consequence, extra-variation is typically observed, and the phenomenon is particularly reinforced by the limited number of replicates and high-dimensionality.

The multinomial distribution is the natural probabilistic model to describe multivariate count data but, in presence of overdispersion, it typically leads to nominal variances well below to the empirical variability. It is possible to cope with overdispersion by several strategies.

Quasi-likelihood assumes that the variance depends on a dispersion parameter $\phi$ representing overdispersion \cite{iii1995} and instead of defining a probabilistic form for the distribution of the data it is sufficient to specify only the variance-mean relationship. A specific quasi-likelihood approach for multivariate count data was investigated by \cite{morel1999covariance}. In the recent years, some generalizations have been proposed: \cite{afroz2020estimating} developed an alternative way of estimating $\phi$ when data are sparse, while \cite{alonso2017new} explained how to deal with clustered multinomial data and unequal cluster size.

Despite the quasi-likelihood approach is robust and works well with severe overdispersion, the problem can be also dealt with alternative and extended family of distributions in a maximum-likelihood perspective. Among these, the Dirichlet-Multinomial compound model \cite{mosimann1962compound} represents one of the most common solutions, able to capture extra-variability by a simple prior on the multinomial parameters. The study and the comparison of the main probabilistic models of the statistical literature able to capture extra-variation in multivariate count data are the focus of this work. A new model that extends the Dirichlet-Multinomial in a deep fashion is also presented together with its parametric-estimation procedure. More precisely the model resembles the deep learning architecture composed by an additional hidden layer with several nodes \cite{Schmidhuber2015}.
A relevant aspect of this model is that the simulation study suggests that its variance tends to the computed variance when the number of nodes goes to $+\infty$.

The rest of this paper is organized in the following way.
In Section~\ref{sec:overdispersion} we present the main parametric models of the statistical literature accounting for extra-variation. We will examine in depth the approaches that are adequate to deal with high-dimensional data. The new model and its estimation procedure are introduced in Section~\ref{sec:new_model}. A simulation study showing the empirical performance of the different strategies is presented in Section~\ref{sec:simulation}. Conclusions and final remarks can be found in Section~\ref{sec:final}.

\section{Models for overdispersion}
\label{sec:overdispersion}
Let \textbf{Y} $= (Y_{1}, \ldots, Y_{j}, \dots, Y_{p})$ be a multivariate vector of counts, where $p$ denotes the total number of categories.
In the Multinomial distribution
\begin{equation}
	P(\textbf{Y} = \textbf{y}) = \frac{m!}{y_{1}! \dots y_{p}!} \prod_{j=1}^p \pi_{j}^{y_{j}}
\end{equation}
where $\bm{\pi} = (\pi_1, \dots, \pi_p)^{T}$ represents the success probability of each of the $p$ categories with $0 \le \pi_j \le 1$ and $\sum_{j=1}^{p} \pi_j = 1$ and $m = \sum_{j=1}^{p} y_j$ is the size indicating the total number of independent trials. The mean and the variance of the distribution depend on $\bm{\pi}$ and $m$ through
\begin{eqnarray}
	E[\textbf{Y}] &=& m\bm{\pi} \label{eq: ExpMult}\\
	Var[\textbf{Y}] &=& m\{diag(\bm{\pi}) - \bm{\pi}\bm{\pi}^T\} \label{eq: VarMult}.
\end{eqnarray}
The multinomial distribution naturally describes the outcomes of $m$ independent trials into $p$ categories, but in many practical situations the assumption of independence of the trials is not respected resulting in the phenomenon of extra-multinomial variation, as shown in \cite{mosimann1962compound}.
Another aspect of this distribution is that it models negative correlations between categories, as clear by taking the marginals in (\ref{eq: VarMult}) that are $Covar[Y_j,Y_{j'}]=-m \pi_j \pi_{j'}$.

Alternative parametric extra-variation models have been proposed in the literature; they may be distinguished by the reason behind the lack of independence.

\paragraph{Dirichlet-Multinomial}
The first parametric alternative to the multinomial distribution was derived by \cite{mosimann1962compound}, under the assumption that the multinomial probability parameters $\pi_1,..., \pi_p$ are distributed according to a Dirichlet distribution. Since it is the natural conjugate of the multinomial, the resulting compound distribution has a closed form and it takes the name of Dirichlet-Multinomial (DM). It is also known in the statistical literature as Multivariate P\'{o}lya distribution, it being the multivariate version of the Beta-Binomial distribution.

From the compound of the Dirichlet distribution with the Multinomial, the random vector \textbf{Y}$\sim DM_p(\bm{\theta}, m)$ has probability function:
\begin{equation}
	P(\textbf{Y} = \textbf{y}) = \notag\\ \frac{\Gamma( \theta_0)\Gamma(m+1)}{\Gamma(m+ \theta_0)} \prod_{j=1}^p \frac{\Gamma(y_j + \theta_j)}{\Gamma(\theta_j)\Gamma(y_j+1)}
\end{equation}

\noindent where $\theta_0 = \sum_{j=1}^p \theta_j$ and $\Gamma$ is the gamma function. By denoting with $\bm \pi =(\frac{\theta_1}{\theta_0},\ldots,\frac{\theta_p}{\theta_0})$ it is possible to show that the expectation is
\begin{eqnarray}
	E[\textbf{Y}] &=& m\bm{\pi} \label{eq: ExpDM}
\end{eqnarray}
so that it has the same expression of the Multinomial expectation in \eqref{eq: ExpMult}.
The variance is corrected by a term in order to account for the extra-variation of the data
\begin{equation}
	Var[\bm{Y}] = m\{1+\rho^2(m-1)\}\{diag(\bm{\pi})-\bm{\pi}\bm{\pi}'\} \label{eq: VarDM}
\end{equation}
where $\rho$ is the overdispersion parameter defined through $\rho^2 = \frac{1}{1+\theta_0}$, so that $0 < \rho < 1$.
The constant $1+\rho^2(m-1)$ inflates the variance of the multinomial distribution and this is what makes the DM a good distribution for modeling overdispersion.
Notice that when $\rho=0$ the DM distribution coincides with the multinomial one. Having the same kernel form of the multinomial distribution, it is easy to check that the correlations among variables are negative.

A recent extension of the DM has been proposed by \cite{valle2010new}. In this framework the multivariate beta distribution of \cite{olkin2003bivariate} is proposed as prior, resulting in a very flexible model. The model is estimated via an independent Metropolis-Hastings algorithm that makes the fitting computationally demanding as the number of replicates and categories increase.

\paragraph{Random-Clumped Multinomial}
The Random-Clumped Multinomial (RCM) was proposed by \cite{morel1993finite} as an alternative to the Dirichlet-Multinomial distribution with the idea to describe the extra-multinomial variation when the lack of independence is introduced by correlation or clumped multinomial sampling. In RCM the vector of counts \textbf{Y} originates by two parts: the first one takes into account the possibility that in cluster sampling within the cluster there are some identical responses due to individuals that greatly influence each other; the second part considers the remaining independent responses. Formally:

%Let $X, X_1^0, \dots, X_m^0$ be IID random variables following a $M_p(\pi,1)$ distribution. Let $U_1, \dots, U_m$ be IID Uniform(0,1) random variables and consider the overdispersion parameter $\rho$, where $0<\rho<1$. Then for we can define $X_j$ as:
%\begin{equation}
% X_j = \begin{dcases*}
%        X  & when $U_j \le \rho$ \\
%        X_j^0 & when $U_j > \rho$
%        \end{dcases*}
%        j=1, \dots, m
%\end{equation}
%Now it is possible to introduce the random variable $\bm{Y} = \sum_{j=1}^m X_j$ that has the same mean and variance of the Dirichlet Multinomial distribution defined in \eqref{eq: ExpMult} and \eqref{eq: VarDM} respectively. It can be also represented through the following equation:
\begin{equation}
	\bm{Y} = \bm{X}N + (\bm{Z}\mid N) \label{eq: Morel93_1}
\end{equation}
where $\textbf{X}$ is distributed as a multinomial with size 1 and $p$ categories, say $M_p(\bm\pi, 1)$, independently from $N \sim M_2(\rho, m)$, which has a binomial distribution. In the second term, $(\bm{Z}\mid N) \sim M_p(\bm \pi, m-N)$ if $N<m$. The random number of counts $N$ is added to $\textbf{X}$ meaning that the addend $\textbf{X}N$ replicates $N$ times the response given by $\textbf{X}$, whereas $(\bm{Z}\mid N)$ considers the independent responses.

\noindent It is possible to prove that the probability distribution of $\textbf{Y}$ is a finite mixture of multinomials \cite{morel1993finite} and more precisely:
\begin{equation}
	P(\bm{Y}=\bm{y}) =\sum_{j=1}^p \pi_j P(W_j = \bm{y}) \label{eq: Morel93_2}
\end{equation}
where $W_j$ for $j=1, \dots, p-1$ is distributed according to a $M_p((1-\rho)\bm\pi + \rho e_j; m)$ and $W_p \sim M_p((1-\rho)\bm\pi; m)$, $e_j$ is the j-th column of the $(p-1) \times (p-1)$ identity matrix and $\bm{\pi} = (\pi_1, \dots, \pi_p)'$ is a probability vector used both as weights for the mixture and as parameters of the multinomials considered in the mixture itself. The RCM distribution has the same mean and variance of the DM distribution, therefore theoretically speaking it can describe the same amount of extra-variation. Empirical differences are thus only ascribed to the estimation method.

The original model proposed by \cite{morel1993finite} accounts for a single clumping only; \cite{banerjee1999miscellanea} developed an extension which integrate multiple random clumping. Specifically they show that the extended finite mixture distribution is a multinomial mixing distribution with different mixing coefficients. This allows to introduce more flexibility but at the cost of additional complexity.

 The model can be estimated through the Fisher's scoring Method. \cite{neerchal2005improved} proposed a two-stage procedure in computing the maximum likelihood estimates in which at first the algorithm uses theoretical limiting results until convergence and then in the second step an extra iteration with the exact Fisher information matrix is implemented.
The resulting algorithm is less computationally expensive with respect to a simple Fisher scoring algorithm, and, at the same time, leads to a better accuracy. As alternative solution, \cite{raim2014method} developed a very fast estimation procedure based on an hybrid approach. At first an approximation of the Fisher scoring algorithm is considered; after an initial warm-up, the classical Fisher's scoring algorithm is applied. More recently, a minorization-maximization algorithm for fitting the RCM has been proposed by \cite{bregu2021mixture}.

\paragraph{Negative Multinomial}
In the multinomial distribution it is well known that the marginals are binomial variates exhibiting a negative correlation. The same negative association between variables is inherited by the DM and the RCM distributions. The Negative Multinomial (NM) distribution assumes instead a positive correlation between variables. It is simply a generalization of the Negative Binomial when multiple outcomes are considered \cite{zhang2017regression}. In the NM, $\bm{Y}$ has parameters $(\bm{\pi}, \beta)=(\pi_1, \dots, \pi_{p+1}, \beta)$, $\sum_{j=1}^{p+1}\pi_j=1$, $\beta>0$ and the probability mass function is defined as
\begin{align}
P(\bm{Y}=\bm{y}) = \binom{\beta+m-1}{m}\binom{m}{\bm{y}}\prod_{j=1}^p \pi_j^{y_j}\pi_{p+1}^{\beta}, \label{eq: NM}
\end{align}
where $m$ is the size and $\pi_{p+1}=1-\sum_{j=1}^p \pi_j$ is the probability of a failure. The first two moments of this distribution are the following
\begin{eqnarray}
	E[\textbf{Y}] &=& \beta\frac{\bm{\pi}}{\pi_{p+1}} \label{eq: ExpNM}\\
	Var[\textbf{Y}] &=& \frac{\beta}{\pi_{p+1}^2}\bm{\pi}\bm{\pi}'+\frac{\beta}{\pi_{p+1}}diag(\bm \pi) \label{eq: VarNM}
\end{eqnarray}
The model can be fitted via maximum likelihood by an iteratively reweighted Poisson regression (see \cite{zhang2017regression} and \cite{manualR2017Zhang} for further details).

\paragraph{Generalized Dirichlet Multinomial}
The Generalized Dirichlet-Multinomial (GDM) was proposed by \cite{connor1969concepts} with the aim to have a general covariance matrix and correlation structure among variables.
The basic idea is to choose a more flexible mixing distribution as a prior for the multinomial given by a kind of generalized Dirichlet distribution.
Following the notation of \cite{zhang2017regression}, the probability mass function of the GDM is
\begin{eqnarray}
	P(\bm{Y}=\bm{y}) = \frac{m!}{y_{1}! \dots y_{p}!}\prod_{j=1}^{p-1} \frac{\Gamma(\alpha_j+y_j)}{\Gamma(\alpha_j)} \frac{\Gamma(\beta_j+\sum_{h=j}^ky_h)}{\Gamma(\beta_j)} \frac{\Gamma(\alpha_j+\beta_j)}{\Gamma(\alpha_j+\beta_j+\sum_{h=j}^ky_h)}
\end{eqnarray}
where $(\bm{\alpha}, \bm{\beta})=(\alpha_1, \dots, \alpha_{p-1}, \beta_1, \dots, \beta_{p-1})$ are the parameters of this distribution, with $\alpha_j, \beta_j >0$. When $\beta_j=\sum_{h=j+1}^p \alpha_h$ the GDM reduces to the DM distribution.

The distribution has the following expectation and variance
\begin{eqnarray}
 E[Y_j] &=& m \begin{dcases*}
        \frac{\alpha_1}{\alpha_1+\beta_1}  & $j=1$ \\
        \frac{\alpha_j}{\alpha_j+\beta_j}\prod_{h=1}^{j-1}\frac{\beta_h}{\alpha_h+\beta_h} & $j=2,...,p-1$ \\
        \prod_{j=1}^{p-1}\frac{\beta_j}{\alpha_j+\beta_j} & $j=p$
        \end{dcases*} \label{eq: ExpGDM} \\
	Var[Y_j] &=&  m \frac{\alpha_j}{\alpha_j+\beta_j}\prod_{h=1}^{j-1}\frac{\beta_h}{\alpha_h+\beta_h}  \left[(m-1)\prod_{h=1}^{j-1}\frac{\beta_h+1}{\alpha_h+\beta_h+1}\frac{\alpha_j+1}{\alpha_j+\beta_j+1} \right.  \nonumber \\ && \left. - m\prod_{h=1}^{j-1}\frac{\beta_h}{\alpha_h+\beta_h}\frac{\alpha_j}{\alpha_j+\beta_j}+1 \right] \label{eq: VarGDM}
\end{eqnarray}

Thanks to the generalized prior of the GDM distribution it is possible to get both positive and negative pairwise correlations among the marginals. The estimation of this model can be obtained via maximum likelihood with quasi-Newton iterations (see \cite{manualR2017Zhang} for major details).\\

In addition to these important contributions, other models and extensions were introduced over time to deal with overdispersion. Interesting recent works focused on Conway-Maxwell-Multinomial \cite{morris2020conway} and Multiplicative Multinomial model \cite{altham2012multivariate}. Both strategies are very flexible and they allow for both overdispersion and underdispersion but they incur in a heavy computational burden, making them infeasible for high-dimensional data, like textual data or genomic datasets.

\section{Deep Dirichlet-Multinomial}
\label{sec:new_model}
In order to deal with overdispersion, we propose in this section a new model that consists of a special kind of a mixture of Dirichlet-Multinomial distributions with restrictions on the parameters. This model, called Deep Dirichlet-Multinomial (DDM), is derived from the idea of mixture of mixtures of DM developed by \cite{viroli2021deep}.

More precisely, let $DM(\boldsymbol\theta,m)$ be the probability mass function of a Dirichlet-Multinomial with parameters $\boldsymbol\theta$ and size $m$, then the probability distribution of the DDM model is defined as
\begin{align}
	P(\bm{Y}=\bm{y}) = \sum_{k=1}^K w_k DM(\bm{\beta}(1+\bm{\alpha_k}), m), \label{eq: DDM}
\end{align}
where $w_k$ for $k=1,\dots,K$ represent the weights of the mixture under the assumptions that $0<w_k<1$ and $\sum_{k=1}^K w_k=1$. Both $\bm{\beta}>0$ and $-1<\bm{\alpha_k}<1$ are vectors of length $p$ and remarkable is the fact that each $\bm{\alpha}_k$ can be interpreted as a perturbation parameter. It being defined in (-1,1), its role is to perturbe $\bm{\beta}$ and to get a more flexible model that behaves better in case of overdispersion.

In order to get this result, we derive the moment generating function and the first two moments of the distribution. Let $\bm \theta_k=\bm{\beta}(1+\bm{\alpha_k})$, $ \theta_{0k}=\sum_{j=1}^p \beta_j(1+\alpha_{jk})$ and $\bm \pi_k =\frac{\bm \theta_k}{\theta_{0k}}$, then the moment generating function of the mixture is
\begin{eqnarray}
	\phi_Y(t)&=& \sum_{k=1}^K w_k \phi_{x_k}(t) \nonumber \\
	&=& \frac{\Gamma(m+1)\Gamma(\theta_{0k})}{\Gamma(m + \theta_{0k})} D_m(\theta_k, (e^{t_1}, \dots, e^{t_p})) \nonumber \\
	 \textrm{ with } D_m&=&\frac{1}{m} \sum_{u=1}^m \left[\left( \sum_{j=1}^p \theta_{jk} e^{t_j*u} \right)D_{m-u}\right], D_0=1.
\end{eqnarray}

As far as the expectation is concerned, it may be seen as a weighted sum of the expected values of each DM distributions:
\begin{align}
	E[\bm{Y}]&= \sum_{k=1}^Kw_k m \bm\pi_k. \label{eq: ExpDDM}
\end{align}

By using the moment generating function we can derive the second moment and the variance of the DDM distribution. It is not difficult to prove that the variance can be split into two components, the first one is a weighted sum of within variances, the second term is a sort of between variance part. Formally:
\begin{eqnarray}\label{eq: VarDDM}
	Var[\bm{Y}]&=& \sum_{k=1}^K w_k m\{diag(\bm\pi_k)-\bm\pi_k \bm\pi_k'\}(1+\rho_k^2(m-1)) \nonumber \\
	&+& \sum_{k=1}^K w_km^2 \bm\pi_k\bm\pi_k' - m^2 \left(\sum_{k=1}^K  w_k \bm\pi_k \right)\left(\sum_{k=1}^K  w_k \bm\pi_k \right)'
\end{eqnarray}
where $\rho_k^2= 1/(1+\theta_{k0})$. The between variance is an additional addendum that can capture both over- and under-dispersion.
By marginalizing the quantity along two different categories, say $j$ and $j'$, the covariance formula is straightforward:
\begin{eqnarray}
	Covar[Y_j,Y_{j'}]&=& \sum_{k=1}^K w_k \pi_{jk} \pi_{j'k}(1-\rho_k^2)m(m-1) \nonumber \\
	&-& m^2 \left(\sum_{k=1}^K  w_k \pi_{jk} \right)\left(\sum_{k=1}^K  w_k \pi_{j'k} \right).
\end{eqnarray}

The expression can take both positive and negative values denoting that the distribution is able to cope with flexible correlation structures among variables. This is a very important property in practice, since groups of variables could be positively correlated to each other but negatively correlated with other variables. For instance, genes could be co-expressed together or, alternatively, words used together in the same context, but synonymous are negatively correlated. Here, this extreme flexibility is obtained at the price of an high number of parameters to be estimated, which largely increase with $K$.

Another important result that we were able to achieve computationally consists in proving that the variance of this model tends to the empirical variance when the number of element of the mixture $K$ goes to $+\infty$. This is clear by the two simulation studies shown in Subsection~\ref{subsec: asymptotic}.

Table~\ref{table:1} contains a synthetic summary of the main characteristics of the presented distributions for multivariate count data.

\begin{table*}[t!]
\centering
\noindent
\begin{footnotesize}
\begin{tabular}{|l|c|c|}
\hline
 & \emph{Multinomial (MN)} & \emph{Dirichlet-Multinomial (DM)} \\
\hline
Parameters &  $m, \ \bm{\pi}=(\pi_1,\dots,\pi_{p-1})'$ & $m, \ \rho, \ \bm{\theta}=(\theta_1,\dots,\theta_p)'$ \\
 & $\sum_{j=1}^p \pi_j=1$ & $\theta_0=\sum_{j=1}^p \theta_j$, $\theta^2=\frac{1}{1+\theta_0}$, $\bm \pi=\frac{\bm \theta}{\theta_0}$\\
$\sharp$ parameters & $p$ & $p+1$\\
%PMF  & $\frac{m!}{y_{1}! \dots y_{p}!} \prod_{j=1}^p \pi_{j}^{y_{j}}$ &  $\frac{m!}{y_{1}! \dots y_{p}!} \frac{\Gamma(c)}{\Gamma(m+c)} \frac{\prod_{j=1}^p \Gamma(y_j + c\pi_j)}{\prod_{j=1}^p \Gamma(c\pi_j)}$  \\
Expectation &  $m\bm{\pi}$ &  $m\bm{\pi}$ \\
Variance & $m\{diag(\bm{\pi}) - \bm{\pi}\bm{\pi}^T\}$ &  $m\{1+\rho^2(m-1)\}\{diag(\bm{\pi})-\bm{\pi}\bm{\pi}'\}$ \\
Covariance & Negatively correlated &  Negatively correlated \\
\hline
 & \emph{Random-Clumped Mult. (RCM)} & \emph{Negative Multinomial} \\
\hline
Parameters &   $m, \ \rho, \bm{\pi}=(\pi_1,\dots,\pi_{p-1})'$ & $m, \ \beta, \bm{\pi}=(\pi_1,\dots,\pi_{p})'$\\
  & $0<\rho<1$ and $\sum_{j=1}^p \pi_j =1$ & $\beta>0$ and $\pi_{p+1}=1-\sum_{j=1}^p \pi_j $\\
$\sharp$ parameters & $p+1$ & $p+2$\\
Expectation  & \small $m\bm{\pi}$ & $\beta\frac{\bm{\pi}}{\pi_{p+1}}$\\
Variance  &  $m\{1+\rho^2(m-1)\}\{diag(\bm{\pi})-\bm{\pi}\bm{\pi}'\}$ & $\frac{\beta}{\pi_{p+1}^2}\bm{\pi}\bm{\pi}'+\frac{\beta}{\pi_{p+1}}diag(\bm \pi)$ \\
Covariance & Negatively correlated &  Positively correlated \\
\hline
 & \emph{Generalized DM (DGM)} & \emph{Deep DM (DDM)} \\
\hline
Parameters &   $m, \ \bm{\alpha}=(\alpha_1, \dots, \alpha_{p-1})$ & $m, \ (w_1,\ldots,w_{K-1}), \ \sum_{k=1}^Kw_k=1$\\
  & $\bm{\beta}=(\beta_1, \dots, \beta_{p-1})$& $\bm\beta=(\beta_1,\ldots,\beta_p), \ \bm\alpha_k=(\alpha_1,\ldots,\alpha_p)$\\
  & $\alpha_j>0$, $\beta_j>0$ & $w_k>0, \ \beta_j>0, \ -1 < \alpha_j < 1$\\
$\sharp$ parameters & $2p-1$ & $p(K+1)+K$\\
Expectation  & see equation (\ref{eq: ExpGDM}) & $\sum_{k=1}^Kw_k m \bm\pi_k$ \\
Variance  &  see equation (\ref{eq: VarGDM}) & see equation (\ref{eq: VarDDM}) \\
Covariance & General correlation &  General correlation \\
\hline
\end{tabular}
\caption{Models for multivariate count data}
\label{table:1}
\end{footnotesize}
\end{table*}

\subsection{Model estimation}
\label{subsec:estimation}
Given a set of observations $(\bm y_1,\ldots,\bm y_n)$ under the assumption of IID random variables, the log-likelihood of the model can be written as

\begin{eqnarray}\label{eq: EM}
  \ell(\bm \Theta) &=& \sum_{i=1}^n \log \sum_{k=1}^K w_k DM(\bm \beta(1+\bm \alpha_k),m) \nonumber \\
  &=& \sum_{i=1}^n \log \sum_{k=1}^K w_k \frac{\Gamma(\theta_{0k})\Gamma(m+1)}{\Gamma(\theta_{0k}+m)} \prod_{j=1}^p \frac{\Gamma(y_{ij}+\theta_{jk})}{\Gamma(\theta_{jk})\Gamma(y_{ij}+1)}
\end{eqnarray}

where $\bm \Theta$ denotes the full set of parameters, and, as defined before, $\bm \theta_k=\bm \beta(1+\bm \alpha_k)$ and $\theta_{0k}=\sum_{j=1}^p \theta_{jk}$.

Parameters in (\ref{eq: EM}) can be efficiently estimated through a generalized EM algorithm \cite{Dempster1977} with a quasi-Newton optimization step for $\bm \beta$ and $\bm \alpha_k$. The EM algorithm maximizes the conditional expectation of the so-called complete
density given the observable data and alternates between the
expectation and the maximization steps until convergence. Let $z$ be the allocation variable of the mixture model defined in (\ref{eq: DDM}) denoting the component membership of each observation. By definition $z$ follows a multinomial distribution
\begin{eqnarray*}
f(z|\boldsymbol\Theta)= \prod_{k=1}^Kw_k^{z_k},
\end{eqnarray*}
from which $f(z_k=1|\boldsymbol\Theta)= w_k$. Evidently, the conditional density of each $\bm y_i$, given the allocation variable, is the $k$th DM distribution.

Then the parameter function to be maximized is the conditional expectation of the
complete density $f(\bm y,z|\boldsymbol\Theta)$ given the observable data, using a fixed set of parameters $\boldsymbol\Theta'$:
\begin{eqnarray}\label{eq: lik}
&& \arg \max_{\boldsymbol\Theta}E_{z|\bm y; \boldsymbol\Theta'}\left[ \log
f\left(\bm y,z|\boldsymbol\Theta \right) \right] \\ \nonumber &=&\arg \max_{\boldsymbol\Theta}E_{z|\bm y;
\boldsymbol\Theta'}\left[ \log f\left(\bm y|z;\boldsymbol\Theta \right) + \log
f\left(z|\boldsymbol\Theta \right)\right].
\end{eqnarray}
By observing
\begin{eqnarray*}
f(\bm y | z,\boldsymbol\Theta)= \prod_{k=1}^K DM(\bm y_i;\bm \beta(1+\bm \alpha_k),m)^{z_k},
\end{eqnarray*}
it is easy to see that formula (\ref{eq: lik}) is equivalent to maximizing the following function with respect to $\boldsymbol\Theta$:
\begin{eqnarray}\label{eq: lik2}
L \left(\boldsymbol\Theta\right)&=& \sum_{i=1}^n \sum_{k=1}^K\tau_{ik} \log \left[ w_k DM(\bm y_i;\bm \beta(1+\bm \alpha_k),m)\right] \nonumber \\
&&  \sum_{i=1}^n \sum_{k=1}^K \tau_{ik} \log w_k +  \sum_{i=1}^n \sum_{k=1}^K \tau_{ik} \log DM(\bm y_i;\bm \beta(1+\bm \alpha_k),m)
\end{eqnarray}

where $\tau_{ik}$ is the posterior
probability that $\bm y_i$ belongs to the
$k$th component of the mixture:
\begin{equation}\label{eq: Estep}
\tau_{ik}=\frac{w_k DM(\bm y_i;\bm \beta(1+\bm \alpha_k),m)}
{\sum_{h=1}^K w_h DM(\bm y_i;\bm \beta(1+\bm \alpha_h),m)}.
\end{equation}

\noindent At each iteration, in the E-step we compute the posterior distributions $\tau_{ik}$ as function of the current set of parameters.
In the M-step we separately maximize the two terms in (\ref{eq: lik2}) under the parameter constraints.

The estimation of the mixture weights under the constraints that they are positive and sum to one is in closed form and takes the usual formula of mixture models:
\begin{eqnarray}\label{eq: Mstep1}
\hat{w}_k=\frac{\sum_{i=1}^n\tau_{ik}}{n}.
\end{eqnarray}

Maximization of the positive vectors $\boldsymbol\beta$ and constraint vectors $\boldsymbol\alpha_k$ involves the derivative of $\log P\left(\bm y_{i} | z_{i}=k; \boldsymbol\Theta\right)$ that can be rewritten as

\begin{eqnarray*}
\log P(\bm y_{i} | z_i= k; \boldsymbol\Theta) &\propto&  \log \Gamma \left(\sum_{j=1}^p \theta_{jk} \right) -  \log \Gamma \left(\sum_{j=1}^p y_{ij} + \theta_{jk} \right) \\
&-& \sum_{j=1}^p \log \Gamma \left( \theta_{jk} \right) +  \sum_{j=1}^p \log \Gamma \left( y_{ij} + \theta_{jk} \right).
\end{eqnarray*}

By remembering $\theta_{jk}=\beta_j(1+\alpha_{jk})$, the gradient of the previous term with respect to the vectors $\boldsymbol \beta$ and $\boldsymbol \alpha_k$ can be obtained as function of digamma defined as $\psi (x) = \frac{d}{dx} \log \Gamma(x)$. Let $ \textbf{1}$ be a column vector of ones of length $p$. The score with respect to $\bm \beta$ is
\begin{eqnarray*}
&&\frac{\partial \log P(\bm y_{i} | z_i= k; \boldsymbol\Theta)}{\partial \bm \beta} =  S_k(\bm \beta)=\psi \left (\boldsymbol\theta_k^\top \textbf{1} \right) (1 + \bm \alpha_k^\top) \\ && - \psi \left (\sum_{j=1}^p y_{ij} + \theta_{jk} \right) (1+\bm \alpha_k^\top)
 - \left(\psi(\theta_{1k})(1+\alpha_{1k}),\ldots,\psi(\theta_{pk})(1+\alpha_{pk}) \right)\\
&&+\left(\psi(\theta_{1k}+y_{i1})(1+\alpha_{1k}),\ldots,\psi(\theta_{pk}+y_{ip})(1+\alpha_{pk}) \right).
\end{eqnarray*}

\noindent Similarly, the score with respect to $\bm \alpha_k$ is
\begin{eqnarray*}
&&\frac{\partial \log P(\bm y_{i} | z_i= k; \boldsymbol\Theta)}{\partial \bm \alpha_k} =  S_k(\bm \alpha_k)=\psi \left (\boldsymbol\theta_k^\top \textbf{1} \right) \bm \beta_k^\top \\ && -
 \psi \left (\sum_{j=1}^p y_{ij} + \theta_{jk} \right) \bm \beta^\top - \left(\psi(\theta_{1k})\beta_{1},\ldots,\psi(\theta_{pk})\beta_{p} \right)\\
&&+\left(\psi(\theta_{1k}+y_{i1})\beta_{1},\ldots,\psi(\theta_{pk}+y_{ip})\beta_{p} \right).
\end{eqnarray*}
Given these scores it is evident that no solution exists in closed form. However at each iteration of the EM algorithm, estimates can be obtained according to quasi-Newton strategies. The scheme of the algorithm is the following:

\bigskip

\bigskip

\bigskip

\noindent\rule[0.5ex]{\linewidth}{1pt}
\begin{enumerate}
  \item \emph{Initialization}: Set $h=0$. For each component $k=1,\ldots,K$, choose values for the vectors $\boldsymbol\alpha_k^{(h)}$ and $\boldsymbol\beta^{(h)}$ and fix equispaced probabilities for $w_k^{(h)}$.
  \item \emph{Estimation step}: Repeat the following until $\ell(\boldsymbol\Theta)$ stops changing:
  \begin{enumerate}
    \item Compute the posteriors using (\ref{eq: Estep});
    \item For $k=1,\ldots,K$ compute new values for $\boldsymbol\alpha _k$ using the scores $S_k(\bm \alpha_k)$ by constrained quasi-Newton.
    \item Compute new values for $\boldsymbol\beta$ using the weighted sum of scores $\sum_{k=1}^K w_k S_k(\bm \beta)$ by constrained quasi-Newton.
    \item For $k=1,\ldots,K$ compute new values for $w_k$ using (\ref{eq: Mstep1}).
    \item $h=h+1$.
  \end{enumerate}
\end{enumerate}
\noindent\rule[0.5ex]{\linewidth}{1pt}

The algorithm has been implemented in R code and is available upon request.

\section{Empirical results}
\label{sec:simulation}

\subsection{Performance comparison}
\label{subsec: comparison}
We illustrate the utility and the properties of the proposed Deep Dirichlet-Multinomial model through two simulation studies. The two empirical studies differ in the way the overdispersion is introduced. More specifically, an increasing percentage of zeros is introduced into the data in order to gradually check the capability of the different probabilistic models to deal with overdispersion.
Data are first randomly generated by a multinomial distribution. Then, in the first scenario, the zeros are added in a completely random way into the dataset. In the second scenario, we added the zeros by gradually replacing the smallest counts, starting from cells with frequency one, ending up to larger counts.

We make a comprehensive comparison of the likelihood-based models discussed in this work and summarized in Table~\ref{table:1}. To this aim, we generated 100 datasets with 10 levels of increasing overdispersion. The ten levels correspond to an increasing proportion of zeros, through jumps of $10\%$, starting from the case of lack of extra-variation with a percentage of added zeros equal to $0\%$ to the case of maximum overdispersion of the data with a percentage of added zeros equal to $90\%$.

The models are fitted on each dataset of the two empirical studies considering datasets with different combinations of samples $n$ and categories $p$ randomly generated from a multinomial distribution with parameters $m=100$ and $\bm{\pi} \sim Unif[0,1]$ then normalized. Here we present the results for $n=50$ rows and $p=20$ columns. With respect to the DDM distribution, we considered three cases each one differentiated by the number of mixture components. More precisely, we estimated the DDM model with $K=2$ and $K=20$ components and, in addition, we also considered the number of components for which the Akaike Information Criterion (AIC) \cite{akaike1974new} is minimized, that is $K=3$ in the first simulation and $K=4$ in the second one.\\

In Table~\ref{table:2} the analytical results of this comparison are shown. In particular, it reports the Euclidean distances of the estimated variance of each model with respect to the empirical one, the mean of the Bayesian Information Criterion (BIC) \cite{schwarz1978estimating} and the mean of the AIC across the 100 replicated datasets.
According to these results, it is clear that the proposed Deep Dirichlet-Multinomial is the model able to better describe the variability of the data, it having the smallest euclidean distance. However, this is achieved at the price of a large number of parameters. In fact, the two information criteria considered in this simulation are both largely penalized by the number of parameters to be estimated. For instance a model with $K=3$ components involves $83$ parameters with respect to the $20$ parameters of a simple multinomial distribution. As a consequence, the DDM is never suggested by the two information criteria.
Table~\ref{table:3} displays the detailed values of the average AIC and BIC in each of the ten scenario. It is possible to notice that as the amount of extra-variation in the data increases, the DDM model gets better and better with respect to the AIC and BIC criteria, even though they do not consider the DDM model as the best choice.\\

\begin{table*}[t!]
\centering
\noindent
\begin{footnotesize}
\begin{tabular}{|l|ccc|l|ccc|}
\hline
\multicolumn{4}{|c|}{\emph{First simulation}} & \multicolumn{4}{c|}{\emph{Second simulation}}\\
\hline
 & \multirow[t]{2}{*}{Euclidean} & \multirow{2}{*}{BIC} & \multirow{2}{*}{AIC} & & \multirow[t]{2}{*}{Euclidean} & \multirow{2}{*}{BIC} & \multirow{2}{*}{AIC}\\
 & Distance & & & & Distance & & \\
\hline
MN & 6.26 & 4248.42 & 4212.73 & MN & 13.66 & 3668.05 & 3643.57\\
DM & 4.08 & 3113.65 & 3076.07 & DM & 15.54 & 2189.94 & 2163.60 \\
RCM & 3.88 & 4088.13 & 4049.89 & RCM & 8.19 & 3323.21 & 3284.96 \\
NM & 5.75 & 4607.13 & 4567.64 & NM & 12.63 & 4004.88 & 3976.68 \\
GDM & 4.96 & 3671.33 & 3599.10 & GDM & 10.33 & 2676.53 & 2622.93 \\
DDM\_2 & 3.84 & 14358.55 & 14243.89 & DDM\_2 & 14.50 & 12025.18 & 11944.31 \\
DDM\_3 & 3.57 & 14395.17 & 14241.03 & DDM\_4 & 12.72 & 12083.04 & 11945.77 \\
DDM\_20 & 2.19 & 15305.45 & 14480.10 & DDM\_20 & 6.87 & 12706.36 & 12117.93\\
\hline
\end{tabular}
\caption{Euclidean distances between the empirical and the estimated variances, BIC, and AIC in the two simulations}
\label{table:2}
\end{footnotesize}
\end{table*}

\begin{figure*}[h!]
\centering
	\includegraphics[width=140mm]{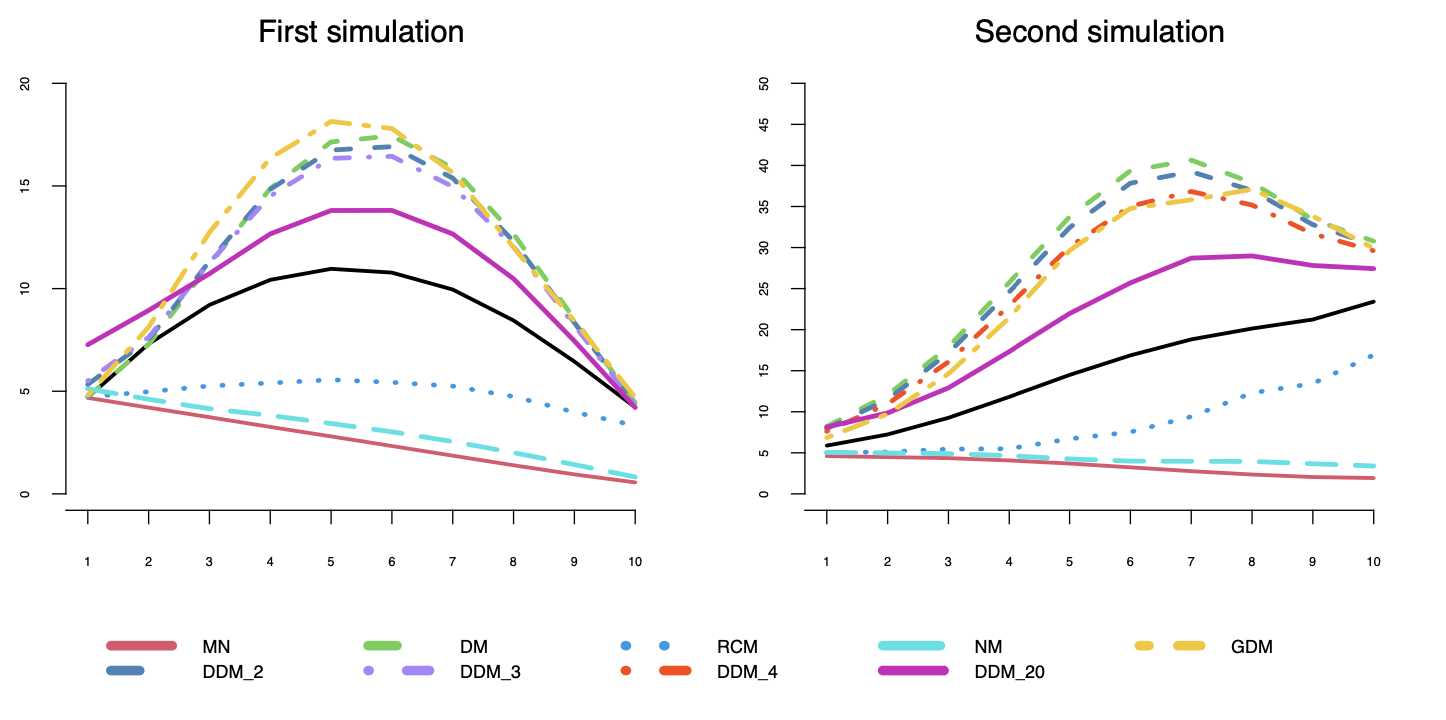}
	\caption{Results of the two simulations. The continuous dark line represents the true empirical variance.}
	\label{fig: Sim1}
\end{figure*}

\begin{table*}[p]
\centering
\noindent
\begin{footnotesize}
\begin{adjustwidth}{-25mm}{-10mm}
\begin{tabular}{|l|cccccccccc|cccccccccc|}
\hline
\multicolumn{11}{|c|}{First simulation}\\
\hline
& \multicolumn{10}{|c|}{\emph{AIC}}\\
\hline
& 1 & 2 & 3 & 4 & 5 & 6 & 7 & 8 & 9 & 10 \\
\hline
MN & 3730.11 & 4239.87 & 4632.08 & 4903.35 & 5028.77 & 4975.53 & 4735.49 & 4253.69 & 3454.69 & 2173.68\\
DM & 3731.78 & 4128.71 & 4214.49 & 4074.22 & 3778.09 & 3364.69 & 2855.69 & 2256.64 & 1567.44 & 788.89\\
RCM & 3731.99 & 4233.06 & 4603.00 & 4899.47 & 4943.83 & 4736.57 & 4351.88 & 3731.40 & 2757.42 & 2510.24\\
NM & 4070.05 & 4604.04 & 5016.75 & 5300.11 & 5427.75 & 5373.01 & 5121.88 & 4615.92 & 3764.81 & 2382.11\\
GDM & 3763.12 & 4113.41 & 4169.15 & 4028.27 & 3739.96 & 3347.01 & 2858.81 & 2276.61 & 1586.73 & 777.39\\
DDM\_2 & 28063.69 & 25071.28 & 21944.59 & 18695.86 & 15497.01 & 12342.16  & 9268.16 & 6359.46 & 3720.23 & 1476.42\\
DDM\_3 & 28084.15 & 25059.62 & 21922.96 & 18677.10 & 15484.21 & 12332.46  & 9264.35 & 6362.05 & 3729.29 & 1494.10\\
DDM\_20 & 28574.69 & 25235.94 & 22009.65 & 18762.76 & 15591.04 & 12474.45 & 9451.56 & 6618.32 & 4080.58 & 2002.07\\
\hline
\hline
& \multicolumn{10}{|c|}{\emph{BIC}}\\
\hline
& 1 & 2 & 3 & 4 & 5 & 6 & 7 & 8 & 9 & 10 \\
\hline
MN & 3766.33 & 4276.08 & 4668.28 & 4939.53 & 5064.91 & 5011.65 & 4771.59 & 4289.71 & 3490.23 & 2205.93\\
DM & 3769.91 & 4166.83 & 4252.59 & 4112.31 & 3816.14 & 3402.72 & 2893.70 & 2294.58 & 1604.87 & 822.89\\
RCM & 3770.23 & 4271.30 & 4641.24 & 4937.71 & 4982.07 & 4774.81 & 4390.13 & 3769.64 & 2795.66 & 2548.48\\
NM & 4110.09 & 4644.08 & 5056.77 & 5340.11 & 5467.71 & 5412.95 & 5161.80 & 4655.77 & 3804.13 & 2417.86\\
GDM & 3835.54 & 4185.83 & 4241.53 & 4100.64 & 3812.24 & 3419.23 & 2930.98 & 2348.70 & 1657.47 & 840.24\\
DDM\_2 & 28179.98 & 25187.57 & 22060.83 & 18812.03 & 15613.07 & 12458.16  & 9384.11 & 6475.17 & 3834.38 & 1580.17\\
DDM\_3 & 28240.48 & 25215.95 & 22079.21 & 18833.27 & 15640.23 & 12488.40  & 9420.22 & 6517.61 & 3882.75 & 1633.59\\
DDM\_20 & 29411.66 & 26072.91 & 22846.22 & 19598.93 & 16426.40 & 13309.41 & 10286.12 & 7451.23 & 4902.37 & 2749.27\\
\hline
\noalign{\smallskip}
\noalign{\smallskip}
\noalign{\smallskip}
\noalign{\smallskip}
\noalign{\smallskip}
\noalign{\smallskip}
\noalign{\smallskip}
\noalign{\smallskip}
\noalign{\smallskip}
\hline
\multicolumn{11}{|c|}{Second simulation}\\
\hline
& \multicolumn{10}{|c|}{\emph{AIC}}\\
\hline
& 1 & 2 & 3 & 4 & 5 & 6 & 7 & 8 & 9 & 10 \\
\hline
MN & 3864.69 & 3985.56 & 4113.77 & 4248.38 & 4282.16 & 4164.47 & 3833.88 & 3324.92 & 2669.02 & 1948.80\\
DM & 3715.84 & 3579.93 & 3329.28 & 2973.99 & 2521.64 & 2009.38 & 1485.57 & 1021.85 & 636.70 & 361.79\\
RCM & 3861.84 & 3946.25 & 4591.91 & 4005.13 & 3909.93 & 3623.74 & 3270.97 & 2559.22 & 1893.32 & 1166.15\\
NM & 4201.49 & 4324.50 & 4455.79 & 4595.27 & 4638.54 & 4527.98 & 4201.10 & 3674.59 & 2969.75 & 2177.78\\
GDM & 3683.84 & 3475.20 & 3259.97 & 2968.15 & 2537.25 & 2043.34 & 1474.48 & 1069.35 & 704.85 & 379.17\\
DDM\_2 & 25407.60 & 22935.99 & 19901.38 & 16401.52 & 12751.42 & 9195.60  & 6079.46 & 3714.90 & 2035.95 & 1019.26\\
DDM\_4 & 25404.66 & 22926.15 & 19887.64 & 16392.59 & 12745.27 & 9193.09 & 6086.79 & 3728.17 & 2053.78 & 1039.58\\
DDM\_20 & 25637.07 & 23075.03 & 19989.91 & 16483.46 & 12858.68 & 9339.59 & 6276.71 & 3941.22 & 2293.48 & 1284.09\\
\hline
\hline
& \multicolumn{10}{|c|}{\emph{BIC}}\\
\hline
& 1 & 2 & 3 & 4 & 5 & 6 & 7 & 8 & 9 & 10 \\
\hline
MN & 3899.09 & 4018.45 & 4144.46 & 4277.08 & 4308.81 & 4188.93 & 3855.68 & 3343.74 & 2684.33 & 1959.90\\
DM & 3752.15 & 3614.73 & 3361.88 & 3004.60 & 2550.20 & 2035.75 & 1509.28 & 1042.55 & 653.79 & 374.43\\
RCM & 3900.08 & 3984.49 & 4630.15 & 4043.37 & 3948.17 & 3661.98 & 3309.21 & 2597.46 & 1931.56 & 1204.39\\
NM & 4239.71 & 4361.21 & 4490.30 & 4627.79 & 4669.02 & 4556.26 & 4226.72 & 3697.17 & 2988.62 & 2191.96\\
GDM & 3752.52 & 3540.55 & 3320.33 & 3024.89 & 2588.49 & 2090.30 & 1516.11 & 1104.54 & 735.38 & 400.29\\
DDM\_2 & 25518.44 & 23042.30 & 20001.09 & 16495.26 & 12839.03 & 9276.62 & 6152.50 & 3778.88 & 2088.99 & 1058.73\\
DDM\_4 & 25591.94 & 23105.88 & 20056.38 & 16551.38 & 12893.84 & 9330.66 & 6211.06 & 3837.31 & 2144.56 & 1107.42\\
DDM\_20 & 26435.89 & 23842.14 & 20710.84 & 17162.63 & 13494.89 & 9929.62 & 6810.91 & 4411.64 & 2686.14 & 1578.88\\
\hline
\end{tabular}
\end{adjustwidth}
\caption{AIC and BIC of the ten scenarios in the two simulated studies.}
\label{table:3}
\end{footnotesize}
\end{table*}

The results of the comparison are shown also from a graphical point of view in Figure~\ref{fig: Sim1} that represents the evolution and the trajectory of the empirical true variance -black line- with respect to the estimated variances of the different considered models when the number of zeros in the dataset increases. In both scenarios it is evident the Deep Dirichlet-Multinomial distribution gets the best approximation and the closest estimate to the real variance in case of over-dispersed data.

In the fist simulation study, where the zeros are increasingly inserted in the data in a random way, the empirical variance has a parabolic behavior. This is due to the fact that the empirical variance increases with the addition of zeros in the data until we get to a situation in which half of the data are zeros in the sixth scenario. Here, it is reached the maximum overdispersion and heterogeneity. Then from the seventh scenario, the extra-variation of the data will start to decrease towards the original level because, as the zeros keep increasing, the dataset will tend to be more homogeneous.
This reasonable behavior is reproduced only by the DM, the GDM and the DDM but it is the latter that is the closest to the real variance when a consistent overdispersion is present.

A parabolic shape is present also in the second simulation, in which the trajectory of the curves is somehow different due to the different method used in order to add the zeros into the dataset. This, however, does not change the fact that the DDM is the model best describing the empirical variance, no matter which scenario is considered.

Another relevant aspect of the DDM distribution can be perceived in Figure~\ref{fig: Sim1}. In both studies as the number of mixture components increases from $2$ to $20$, passing through $3$ in the first simulation and $4$ in the second one, the variance of the model gets closer and closer to the empirical one. This aspect will be further discussed in the next section.

\subsection{DDM Asymptotic behavior}
\label{subsec: asymptotic}
\begin{figure*}[t!]
    \centering
    \begin{subfigure}[t]{0.5\textwidth}
        \centering
        \includegraphics[width=7cm]{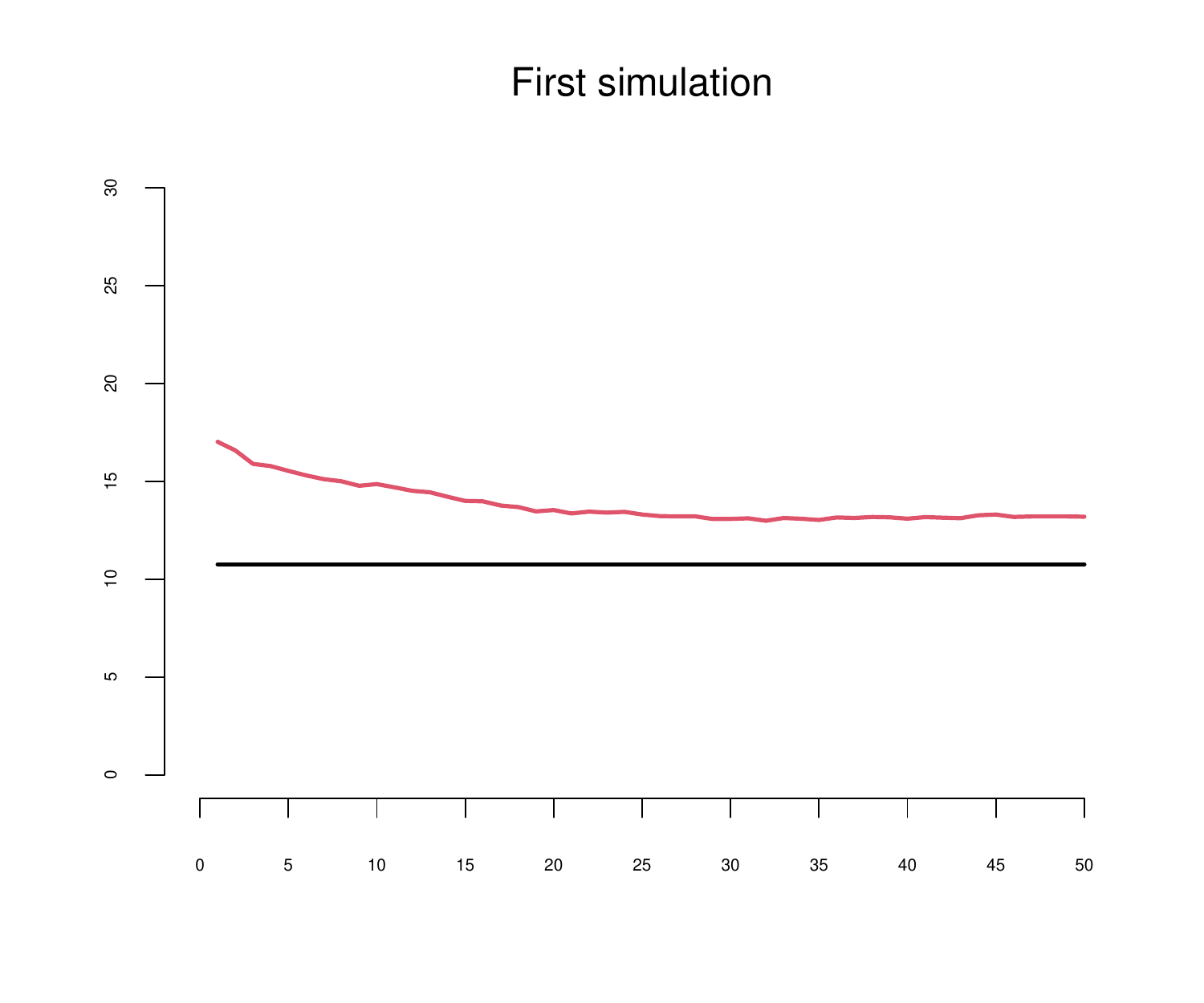}
        \caption{First simulation}
    \end{subfigure}%
    \begin{subfigure}[t]{0.5\textwidth}
        \centering
        \includegraphics[width=7cm]{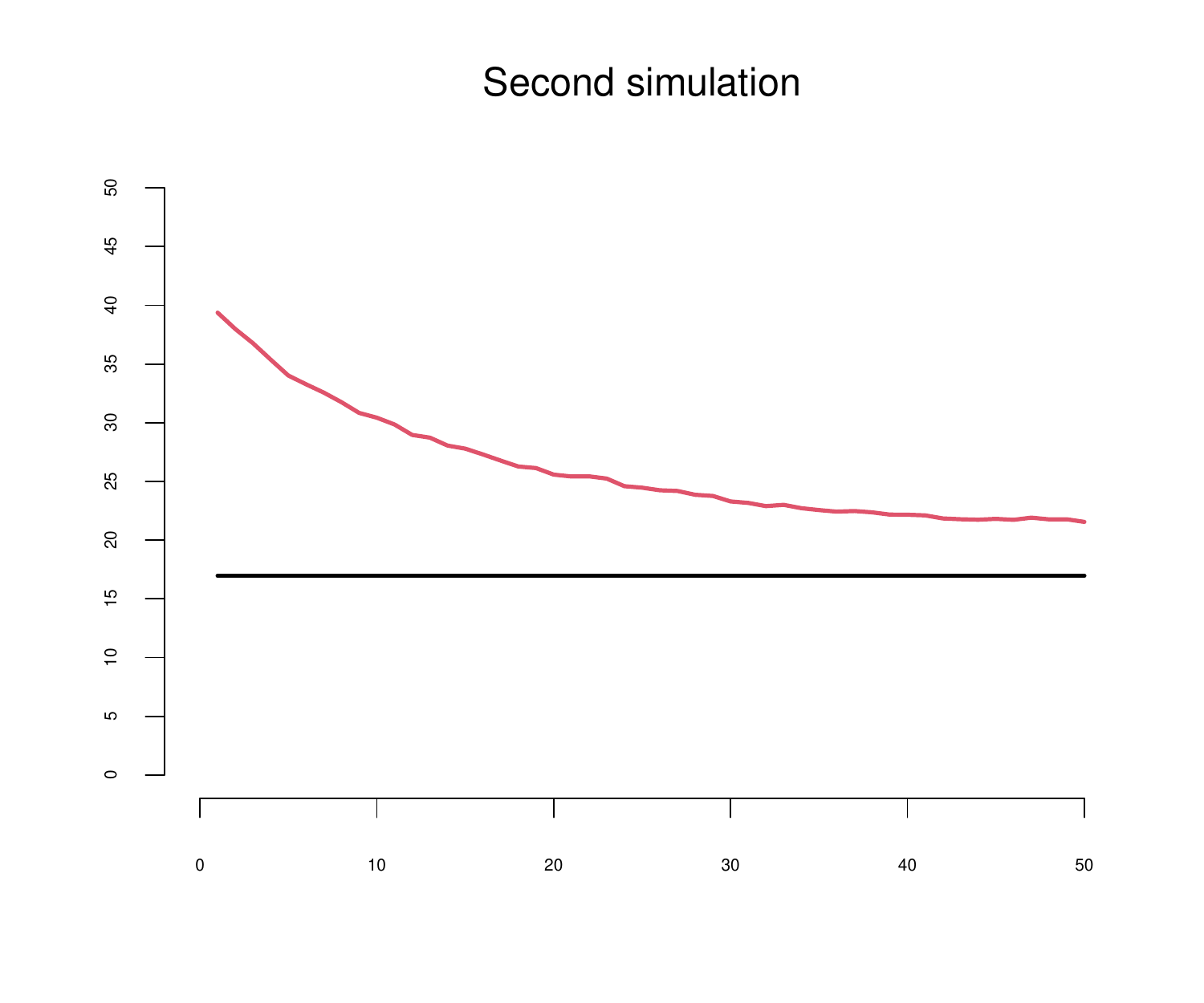}
        \caption{Second simulation}
    \end{subfigure}
    \caption{DDM variance vs. computed variance as $K \rightarrow +\infty$}
    \label{fig: fig2}
\end{figure*}
Exploiting the same data generating process defined for the two simulations above, in order to analyze the asymptotic behavior of the Deep Dirichlet-Multinomial variance we take into consideration only the intermediate scenario with a $50\%$ share of zeros added.

For each value of $K=1,\dots, 50$ the DDM model together with its variance-covariance matrix are estimated with 10 replications each. The summarized results are displayed in Figure~\ref{fig: fig2}. The black line represents the sample variance mean across 10 random datasets, while the red one describes the evolution of the fitted DDM variance when $K$ increases to $+\infty$.
It is straightforward to see how this newly introduced model is well capable of accounting for the extra-variation of the data. In particular, we proved form an empirical point of view that its variance tends to the computed variance when the number of elements K of the mixture go to $+\infty$.

\subsection{Choosing K in DDM distribution}
\label{subsec: choosingK}

In this section we aim at verifying wether the optimal number of mixture components of the DDM distribution can be reasonably suggested by the AIC and BIC indicators.
\begin{figure*}[t!]
\begin{footnotesize}
\centering
\noindent
	\scalebox{.7}{\includegraphics{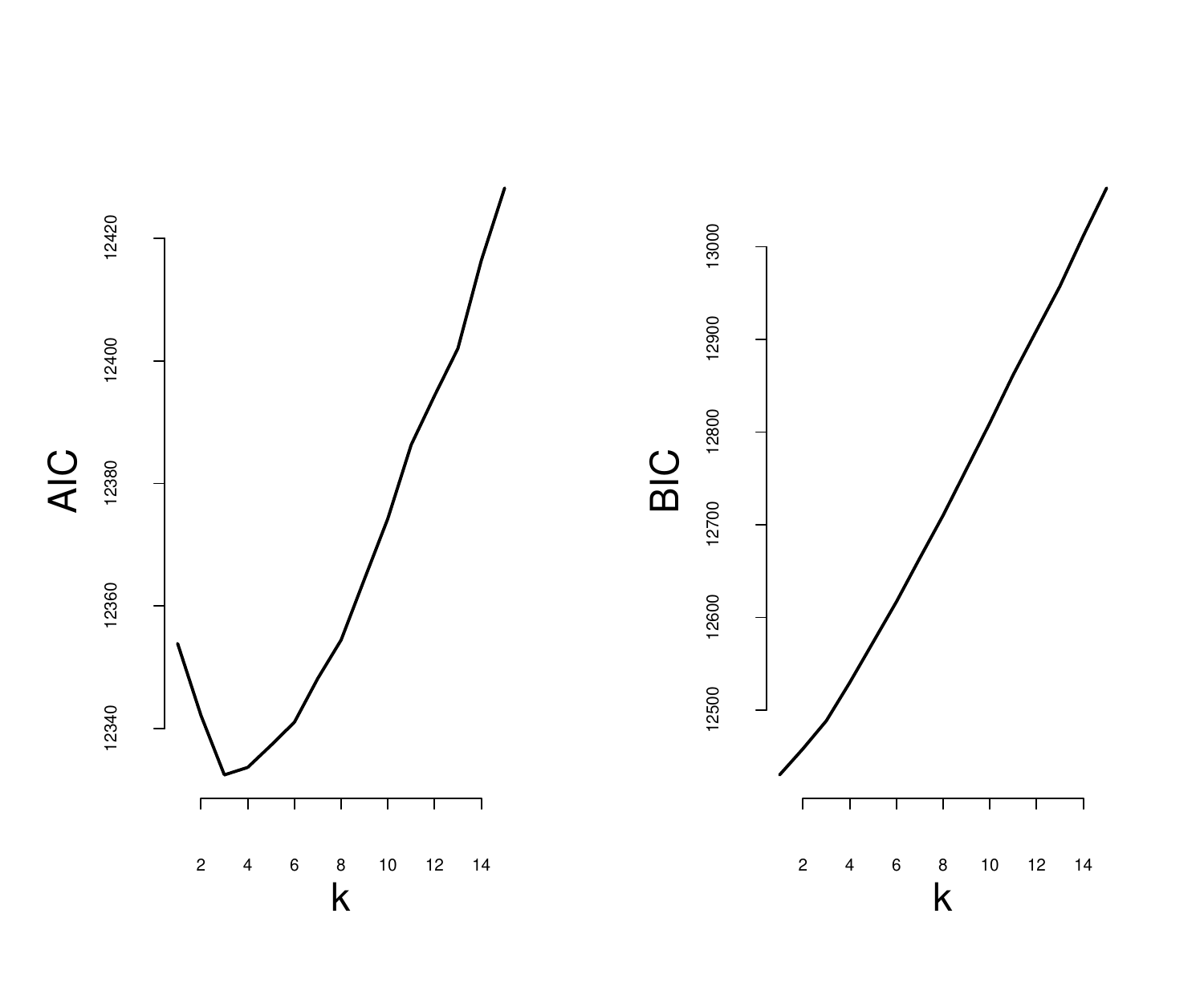}}
	\caption{AIC and BIC for each k in the first simulation}
	\label{fig: AicBic1}
\end{footnotesize}
\end{figure*}
\begin{figure*}[t!]
\begin{footnotesize}
\centering
\noindent
	\scalebox{.7}{\includegraphics{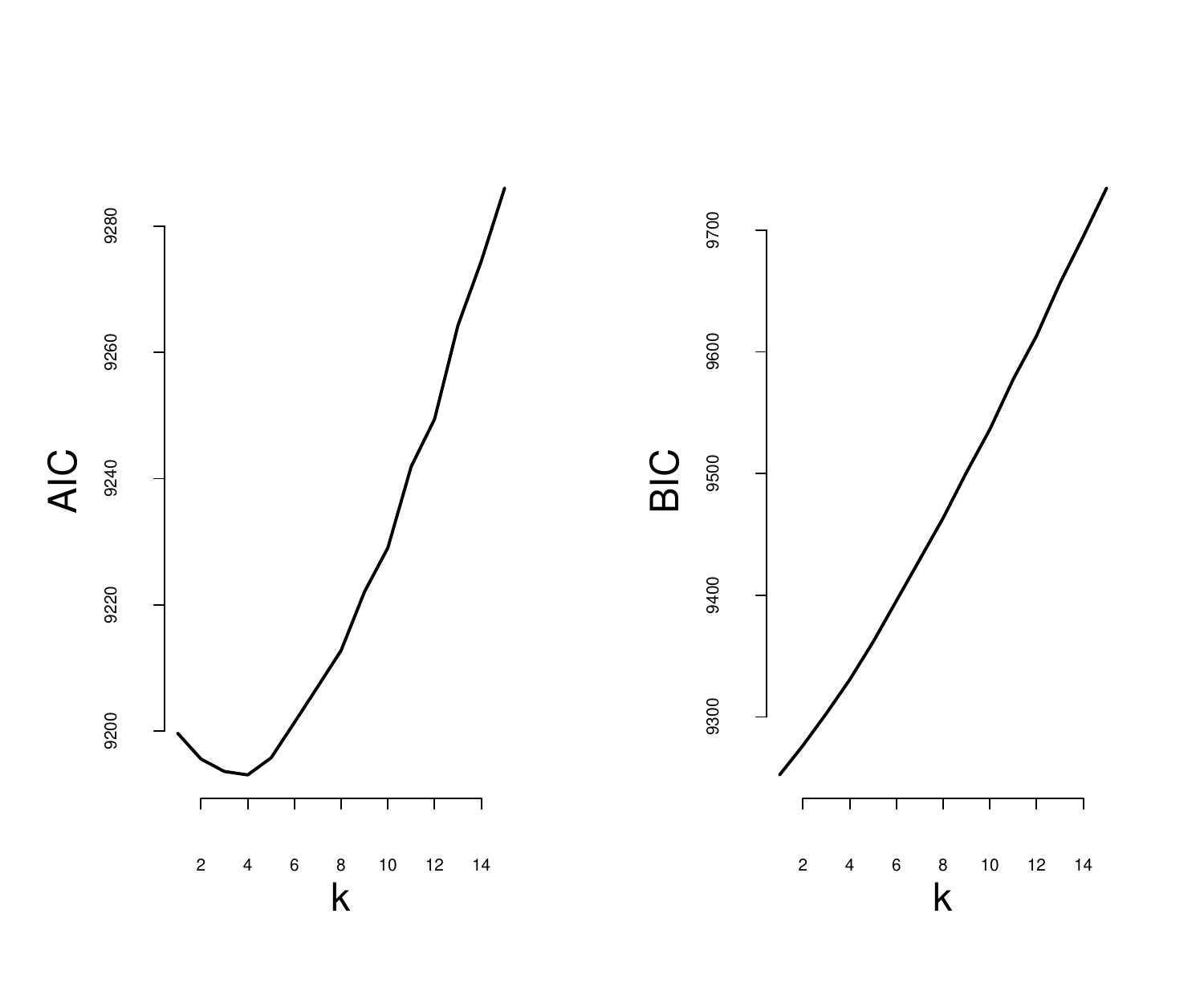}}
	\caption{AIC and BIC for each k in the second simulation}
	\label{fig: AicBic2}
\end{footnotesize}
\end{figure*}

The data were generated from a multinomial distribution as described in Section~\ref{subsec: comparison}, a study for each one of the two simulations defined above was developed and, among the ten different scenarios, we considered the intermediate one which presents a $50\%$ of overdispersion in the data. In particular, the DDM model was estimated 100 times for each value $K=1, \dots, 15$ of mixture components. Figure 3 and 4 shows the AIC and BIC results for each of the two methods of adding the zeros in the dataset.

As expectable in both studies, the BIC is strongly penalized by the number of parameters and presents a strictly increasing trend as the number $K$ of components increases, thus not being able to select an optimal number of components. On the contrary, the AIC, it being less penalized, seems to offer a clear suggestion of the number of components. In this case a minimum is reached at $K=3$ and $K=4$ respectively in the first and second simulation study.

\section{Final remarks}
\label{sec:final}
In this paper we firstly conducted a comprehensive analysis of the likelihood based models able to deal with data that present extra-multinomial variation. We then proposed a new distribution, the Deep Dirichlet-Multinomial, that resembles the deep learning architecture composed by an additional hidden layer with several nodes. Even though the DDM distribution is not always considered to be the best choice by BIC and AIC due to the large amount of parameters that need to be estimated, compared to the other models that were analyzed, it is characterized by some interesting and desirable properties.

First of all, the analytical formula of its variance can be split in two components that ideally represent the within and between variances. This allows to capture both under- and over-dispersion and to have a more flexible correlation structure among variables. Moreover, we showed computationally that the variance of the DDM model tends to the computed variance when the number of mixture components increases, and this is of course a desirable property that a good distribution should have.
The choice of estimating the DDM distribution using an EM algorithm leads to good results in all the simulations considered and permit to stay coherent with the estimation methods of the other models considered in the comparison.

During the comparison of the likelihood-based models for over-dispersed data, we have also considered the idea of zero-inflated models. However, they do not seem to be the best solution when trying to deal with the overdispersion problem in multinomial data. The main reason is due to the fact that in multivariate count data they tend to describe entire rows of zeros, that are instead usually removed when it comes to the analysis of these kind of data.

\bibliographystyle{plain}
\bibliography{References}

\begin{thebibliography}{10}

\bibitem{afroz2020estimating}
Farzana Afroz, Matt Parry, and David Fletcher.
\newblock Estimating overdispersion in sparse multinomial data.
\newblock {\em Biometrics}, 76(3):834--842, 2020.

\bibitem{akaike1974new}
Hirotugu Akaike.
\newblock A new look at the statistical model identification.
\newblock {\em IEEE transactions on automatic control}, 19(6):716--723, 1974.

\bibitem{alonso2017new}
JM~Alonso-Revenga, Nirian Mart{\'\i}n, and Leandro Pardo.
\newblock New improved estimators for overdispersion in models with clustered
  multinomial data and unequal cluster sizes.
\newblock {\em Statistics and Computing}, 27(1):193--217, 2017.

\bibitem{altham2012multivariate}
Pat~ME Altham, Robin~KS Hankin, et~al.
\newblock Multivariate generalizations of the multiplicative binomial
  distribution: Introducing the mm package.
\newblock {\em Journal of Statistical Software}, 46(12):1--23, 2012.

\bibitem{banerjee1999miscellanea}
T~Banerjee and SR~Paul.
\newblock Miscellanea. an extension of morel-nagaraj's finite mixture
  distribution for modelling multinomial clustered data.
\newblock {\em Biometrika}, 86(3):723--727, 1999.

\bibitem{bregu2021mixture}
Ornela Bregu, Nuha Zamzami, and Nizar Bouguila.
\newblock Mixture-based clustering for count data using approximated fisher
  scoring and minorization--maximization approaches.
\newblock {\em Computational Intelligence}, 37(1):596--620, 2021.

\bibitem{connor1969concepts}
Robert~J Connor and James~E Mosimann.
\newblock Concepts of independence for proportions with a generalization of the
  dirichlet distribution.
\newblock {\em Journal of the American Statistical Association},
  64(325):194--206, 1969.

\bibitem{Dempster1977}
A.~P. Dempster, N.~M. Laird, and D.~B. Rubin.
\newblock {Maximum Likelihood from Incomplete Data via the EM Algorithm}.
\newblock {\em Journal of the Royal Statistical Society. Series B
  (Methodological)}, 39(1):1--38, 1977.

\bibitem{efron86}
Bradley Efron.
\newblock Double exponential families and their use in generalized linear
  regression.
\newblock {\em Journal of the American Statistical Association},
  81(395):709--721, 1986.

\bibitem{iii1995}
N.~David~Yanez Iii and Jeffrey~R. Wilson.
\newblock Comparison of quasi-likelihood models for overdispersion.

\bibitem{morel1999covariance}
Jorge~G Morel.
\newblock A covariance matrix that accounts for different degrees of extraneous
  variation in multinomial responses.
\newblock {\em Communications in Statistics-Simulation and Computation},
  28(2):403--413, 1999.

\bibitem{morel1993finite}
Jorge~G Morel and Neerchal~K Nagaraj.
\newblock A finite mixture distribution for modelling multinomial extra
  variation.
\newblock {\em Biometrika}, 80(2):363--371, 1993.

\bibitem{morris2020conway}
Darcy~Steeg Morris, Andrew~M Raim, and Kimberly~F Sellers.
\newblock A conway--maxwell-multinomial distribution for flexible modeling of
  clustered categorical data.
\newblock {\em Journal of Multivariate Analysis}, 179:104651, 2020.

\bibitem{mosimann1962compound}
James~E Mosimann.
\newblock On the compound multinomial distribution, the multivariate
  $\beta$-distribution, and correlations among proportions.
\newblock {\em Biometrika}, 49(1/2):65--82, 1962.

\bibitem{Munzert2015}
Simon Munzert, Christian Rubba, Peter Mei{\ss}ner, and Dominic Nyhuis.
\newblock {\em {Automated Data Collection with R: A Practical Guide to Web
  Scraping and Text Mining}}.
\newblock Wiley, Hoboken, NJ, USA, Jan 2015.

\bibitem{neerchal2005improved}
Nagaraj~K Neerchal and Jorge~G Morel.
\newblock An improved method for the computation of maximum likeliood estimates
  for multinomial overdispersion models.
\newblock {\em Computational Statistics \& Data Analysis}, 49(1):33--43, 2005.

\bibitem{olkin2003bivariate}
Ingram Olkin and Ruixue Liu.
\newblock A bivariate beta distribution.
\newblock {\em Statistics \& Probability Letters}, 62(4):407--412, 2003.

\bibitem{Poortema1999}
K.~Poortema.
\newblock {On modelling overdispersion of counts}.
\newblock {\em Stat. Neerl.}, 53(1):5--20, Mar 1999.

\bibitem{raim2014method}
Andrew~M Raim, Minglei Liu, Nagaraj~K Neerchal, and Jorge~G Morel.
\newblock On the method of approximate fisher scoring for finite mixtures of
  multinomials.
\newblock {\em Statistical Methodology}, 18:115--130, 2014.

\bibitem{Schmidhuber2015}
J{\ifmmode\ddot{u}\else\"{u}\fi}rgen Schmidhuber.
\newblock {Deep learning in neural networks: An overview}.
\newblock {\em Neural Networks}, 61:85--117, Jan 2015.

\bibitem{schwarz1978estimating}
Gideon Schwarz et~al.
\newblock Estimating the dimension of a model.
\newblock {\em Annals of statistics}, 6(2):461--464, 1978.

\bibitem{valle2010new}
Luciana~Dalla Valle and Fabrizio Leisen.
\newblock A new multinomial model and a zero variance estimation.
\newblock {\em Communications in Statistics—Simulation and
  Computation{\textregistered}}, 39(4):846--859, 2010.

\bibitem{viroli2021deep}
Cinzia Viroli and Laura Anderlucci.
\newblock Deep mixtures of unigrams for uncovering topics in textual data.
\newblock {\em Statistics and Computing}, 31(3):1--10, 2021.

\bibitem{Wang2009Jan}
Zhong Wang, Mark Gerstein, and Michael Snyder.
\newblock {RNA-Seq: a revolutionary tool for transcriptomics}.
\newblock {\em Nat. Rev. Genet.}, 10:57--63, Jan 2009.

\bibitem{manualR2017Zhang}
Yiwen Zhang and Hua Zhou.
\newblock {\em {MGLM}: Multivariate Response Generalized Linear Models}, 2018.
\newblock R package version 0.2.0.

\bibitem{zhang2017regression}
Yiwen Zhang, Hua Zhou, Jin Zhou, and Wei Sun.
\newblock Regression models for multivariate count data.
\newblock {\em Journal of Computational and Graphical Statistics}, 26(1):1--13,
  2017.

\end{thebibliography}
\end{document}